# Computation of Madelung Energies for Ionic Crystals of Variable Stoichiometries and Mixed Valencies and their application in Lithium-ion battery voltage modelling

K. Ragavendran, D. Vasudevan, A. Veluchamy and Bosco Emmanuel*

Central Electrochemical Research Institute, Karaikudi-630 006, India

## Abstract

Electrostatic energy (Madelung energy) is a major constituent of the cohesive energy of ionic crystals. Several physicochemical properties of these materials depend on the response of their electrostatic energy to a variety of applied thermal, electrical and mechanical stresses. In the present study, a method has been developed based on Ewald's technique, to compute the electrostatic energy arising from ion-ion interactions in ionic crystals like $Li_xMn_2O_4$ with variable stoichiometries and mixed valencies. An interesting application of this method in computing the voltages of lithium ion batteries employing spinel cathodes is presented for the first time. The advantages of the present method of computation over existing methods are also discussed.

---

* author for correspondence

## Introduction

Electrostatic energy of ionic crystals [1-3] is an important constituent of the cohesive energy [4] of these crystals. Various physicochemical properties such as melting points, heats of fusion and evaporation and activation energies for formation and diffusion of electronic and atomic point defects are related to the solid-state cohesion [5]. Response of the lattice energy towards a variety of applied thermal, electrical and mechanical stresses lead to piezoelectric, ferroelectric and electrochromic properties of these materials. Madelung energy computations have also recently generated much interest [6-9]. A Madelung model has been used to predict the dependence of lattice parameter on the nanocrystal size [6]. Three-dimensional systems periodic in one direction have been simulated using the Ewald summation method [7]. Madelung constants were computed for a wide variety of ionic crystals and it was further shown that structural phase transitions could also be probed within this framework [8]. More recently, Madelung type long-range coulomb interactions were shown to be important in fixing the optimal doping level, i.e., the stoichiometry, in high-temperature superconductors [9].

Electrostatic energies of simple ionic crystals of fixed stoichiometries and valencies like CsCl, NaCl and ZnS have already been calculated and reported as Madelung constants in literature. This energy, however, is no more a constant for non-stoichiometric and multivalent compounds (Eg. $VO_x$, $UO_{2+x}$, $Li_xCoO_2$, $LiMn_2O_4$, $Li_xWO_3$, $Na_xWO_3$, $TiO_x$ and $Li_xMn_{2-y}M_yO_4$ with dopant) as it varies with the stoichiometry as well as with the valency. A method has been developed in the present study to compute the electrostatic energy arising from ion-ion interactions in ionic crystals of variable stoichiometries and mixed valencies. A novel use of this method in computing the voltage of Lithium ion batteries with electrodes of variable stoichiometry and valency is presented with $Li_xMn_2O_4$, a widely studied cathode material used in high-voltage lithium-ion batteries, as a specific example.

**Ewald method applied to ionic crystals of variable stoichiometries and mixed valencies**

In this section, Ewald's technique has been applied to compute the long-range electrostatic interactions in ionic crystals of variable stoichiometries and mixed valencies. Any ionic crystal may be specified by giving its crystallographic space group, the unit cell parameters (corresponding to the primitive, conventional or super cells) and the corresponding basis (consisting of a set of ions). The electrostatic energy of ionic crystals is usually expressed as a sum of pair wise coulombic terms given by

$$E_M = \sum_{(i,j)} \frac{z_i z_j}{r_{ij}} \quad (1)$$

where $z_i$ and $z_j$ are the valencies of the i$^{th}$ and j$^{th}$ ion and $r_{ij}$ is the interionic distance. The sum runs over all ion pairs. In order to apply Ewald's method for crystals of variable stoichiometry and mixed valency, the above sum is expressed in terms of contributions arising from several sublattices present in the crystal so that the stoichiometry and the valency can be tuned in each sublattice. Hence the appropriate form for the energy will be

$$E_M = \frac{1}{2} \sum_{i_{ref}=1}^{N} E_{i_{ref}} \quad (2)$$

where N is the number of ions in the basis and is also the number of sublattices into which the crystal can be split. The factor 1/2 removes the double counting of the pair interaction.

Using Ewalds method, one can obtain the final expression

$$E_M = 1/2[\sum_{g\neq 0} (S^{\cdot}S^{*})\, f(\mathbf{g}) + \{\sum_{i=1}^{N} \lambda^2(i)\}.F(G) + \sum_{j=1}^{N}\lambda(j)\sum_{i\neq j}^{N}\lambda(i)\,\bar{F}(G,\mathbf{r}_i^j)] \qquad (3)$$

where $S = \sum_{i=1}^{N}\lambda(i)\exp ig.ri$

$S^*$ is the complex conjugate of S; G is the convergence factor and $\mathbf{r}_i^j = \mathbf{r}_i - \mathbf{r}_j$.

The details of the derivation and the meaning of symbols appearing in equation (3) are given in the **Appendix.** It must be noted that, though the Ewald method is well known, its present application is new. Equation (3) is the most general form for the electrostatic energy in ionic crystals with variable stoichiometry and mixed valency. It forms the basis of Madelung energy computations for systems with variable stoichiometry and valency.

**Choice of unit cells for crystals with sub-lattice order and sub lattice disorder vis-à-vis variable stoichiometries and valencies**

In the previous section, a method of arriving at a general formula for the electrostatic energy of any ionic crystal with variable stoichiometries and valencies was reported. Before proceeding towards computing the electrostatic energy, it is desirable to clarify the meaning of the $\lambda(i)'$s (i =1 to N).

For regular stoichiometric ionic crystals such as NaCl, CsCl and ZnS, these effective charges are integral quantities. In the case of non-stoichiometric or variable stoichiometric crystals, there arises a need to distinguish between crystals with sublattice order or sublattice disorder. For crystals with sublattice order, these effective charges will again be integral quantities whereas for crystals with sublattice disorder they will be fractional, reflecting the random distribution of ions and their valencies in the sublattices. For the computation, one can choose the primitive unit cell, the conventional unit cell or even the super cell depending on the problem in hand. Choosing primitive unit cell would certainly take the least computational time. However, though it can handle efficiently cases with sublattice disorder, it has obvious limitations for cases with sublattice ordering. Super cells may be required to deal efficiently with cases of sublattice ordering with variable stoichiometry and valence.

In order to clarify the above issues, a discussion using a specific crystal, for examplem, $Li_xMn_2O_4$, will help. This is an oxide belonging to the class of spinels with space group Fd3m [13]. The primitive basis has two lithium ions, eight oxide ions and four manganese ions. The oxide ion valence can be considered fixed at –2 and lithium valence at +1. This crystal is a mixed valent compound with respect to oxidation state of the manganese ion. When the stoichiometry x of the spinel varies from 0 to 1, the valence of the Mn ion continuously varies from a state of all 4+ to a mixed valent state of 50% of 4+ and 50% of 3+. At this point two cases arise: (i) Spinel with sublattice ordering and (ii) Spinel with sublattice disorder. For case (i), the x value is restricted to 0, 0.5 and 1 if we choose the primitive basis for the energy computation and for enlarging the scope to the computation to more values of x one will have to move on from the primitive to conventional and even to super cells. For case (ii), the primitive basis alone can handle all values of x.

**Details of the computation and results**

A programme was written to implement equation 3 for the computation of Madelung energy. Inputs to the programme were as follows: Space group of the spinel, Cubic lattice constant and the atomic positions of the 14 ions in the primitive cell (4 Mn, 8 O and 2 Li ions); Convergence factor G was set as 1 with grid size of 10 x 10 x 10 for both direct and reciprocal lattices; The stoichiometry and valency dependent parameters, viz. the $\lambda(i)'s$, were fixed as follows:

$$\lambda_1 = \lambda_2 = x$$

$$\lambda_3 = \lambda_4 = \lambda_5 = \lambda_6 = 4 - (x / 2)$$

$$\lambda_7 = \lambda_8 \text{--------} = \lambda_{14} = -2$$

It may be further noted from equation 3 that the stoichiometry and valency dependent parameters, viz *the* $\lambda(i)'s$, are product separable from a host of factors, which depend *only* on the crystal structure. These can be viewed as a set of generalised Madelung constants, which take the place of the single Madelung constant for conventional stoichiometric crystals.

Before applying it to cases of variable stoichiometry and valency, equation 3 was tested against conventional systems such as NaCl, CsCl and ZnS. The Madelung energy calculated using equation (3) matched with the values reported in literature [1] correct to five decimal places. When applied to the spinel $Li_xMn_2O_4$, $\lambda(i)'s$ are no more constants as for conventional systems but vary as a function of x. The result of applying equation 3 to compute Madelung energy of this spinel as a function of x is presented in **Fig. 1**. The figure is almost linear.

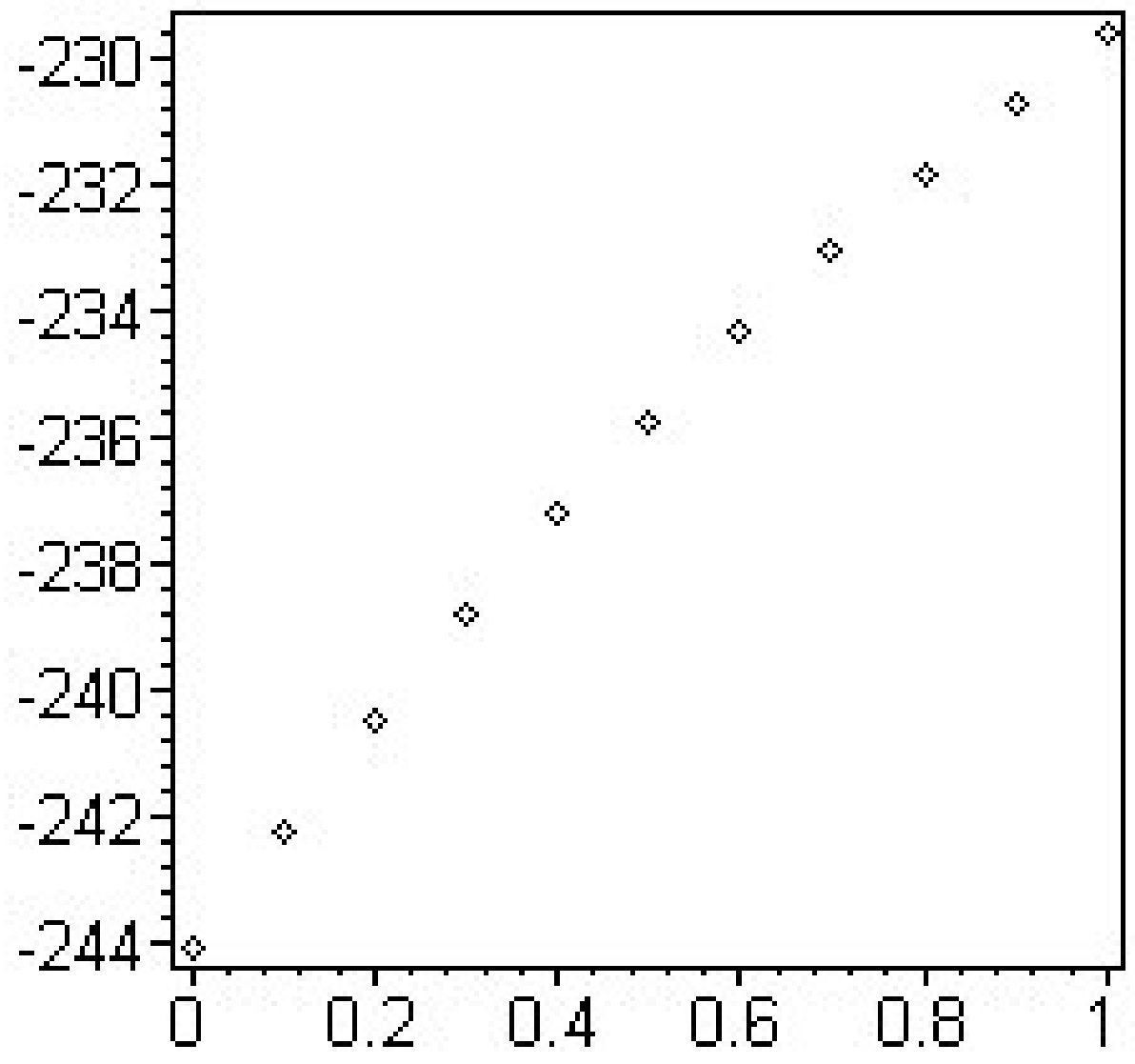

**Figure 1**: Madelung Energy $E_M$ ( in eV/ Formula unit ) Vs. the stoichiometry *x*

The programme was executed on a 1.13 GHz Pentium III and the total energy computation (for all values of x) took nearly eight hours. It is to be noted here that Ceder and coworkers used quantum ab initio methods to compute total energy of layered oxides of lithium on a Cray C 90 Supercomputer, which is reported to take one hour for one total energy calculation [10,11]. The authors have computed total energies for *only x = 0 and x = 1*. For intermediate values of x, one needs to do computation on superstructures, which requires computational speeds that, is beyond the limits of present day resources.

**Discussion and application to lithium-ion battery voltage computation**

The values of $x$=0 and $x$=1 in $Li_xMn_2O_4$ correspond to the fully charged and the fully discharged states of the battery respectively. In this section, the voltage of the battery system is modelled using the Madelung energies computed in the previous sections.

**<u>Battery system</u>: $Li_xMn_2O_4$ / $Li^+$ ion carrying electrolyte / Li metal**

The electrode reactions in this battery can be represented as:

$$\nu\, Li_xMn_2O_4 + Li^+ + e^- \rightarrow \nu\, Li_{x+(1/\nu)}Mn_2O_4 \text{ (at cathode)} \quad \text{(i)}$$

$$Li \rightarrow Li^+ + e^- \text{ (at anode)} \quad \text{(ii)}$$

Adding (i) and (ii),

$$\nu\, Li_xMn_2O_4 + Li \rightarrow \nu\, Li_{x+(1/\nu)}Mn_2O_4 \text{(Overall cell reaction)} \quad \text{(iii)}$$

(Note: For every $\nu$ moles of $Li_xMn_2O_4$ one Faraday passes through the circuit)

The battery voltage is given by

$$V = -\Delta G/F \quad (4)$$

where $\Delta G$ is the free energy change accompanying equation (iii). On neglecting volume and entropy effects [10], equation (4) can be rewritten as $V= -\Delta E/F$, where $\Delta E$ is the energy change of reaction (iii). Before proceeding further, it must be remarked that the energy quantities must be normalised as per equivalent as pointed out by Vijh [12] and Diggle [5]. For example, the energy of formation per mole may be divided by the number of total valencies (either cationic or anionic) participating in the compound to obtain the energy of formation per equivalent [12]. For $Li_xMn_2O_4$, a division by 8 is required. For a convenient evaluation of ΔE for the reaction (iii), the following steps may be considered:

| **Cell Reaction** | **Energy change** | |
|---|---|---|
| $xLi + Mn_2O_4 \rightarrow Li_xMn_2O_4$ | $\overline{\Delta E}\ (x)$ | (iv) |
| $(x + 1/\nu)Li + Mn_2O_4 \rightarrow Li_{(x + 1/\nu)}Mn_2O_4$ | $\overline{\Delta E}\ (x + 1/\nu)$ | (v) |

(v) – (iv) gives

| | | |
|---|---|---|
| $Li_xMn_2O_4 + 1/\nu Li \rightarrow Li_{(x + 1/\nu)}Mn_2O_4$ | $[\overline{\Delta E}\ (x + 1/\nu) - \overline{\Delta E}\ (x)]$ | (vi) |

Thus, $\Delta E = \nu[\overline{\Delta E}\ (x + 1/\nu) - \overline{\Delta E}\ (x)]$

$= [\overline{\Delta E}\ (x + 1/\nu) - \Delta G\ (x)]/ (1/\nu)$

$\Rightarrow d\overline{\Delta E}\ (x)/dx$ (for a differential change in x) (5)

Hence finally, $V = -(1/F)\ d[\overline{\Delta E}\ (x)]/dx]$ (6)

Equation (6) is the desired relation connecting battery voltage V and the energy change $[\overline{\Delta E}\ (x)]$.

$\overline{\Delta E}\ (x)$ can be deduced as follows, by breaking reaction (iv) into elementary Born steps.

| **Born Step** | **Energy change** |
|---|---|
| $x$Li (s)$\rightarrow x$ Li (vap) $\rightarrow x\ Li^+ + x\ e^-$ | $x$ (1.65 + 5.39) eV |
| $Mn_2O_4(s) \rightarrow 2\ Mn^{4+} + 4\ O^{2-}$ (vap) | $E_M(Mn_2O_4)$ |
| $xMn^{4+} + x\ e^- \rightarrow x\ Mn^{3+}$ | -$x$ (52) eV |

$xLi^+ + (2-x)Mn^{4+} + xMn^{3+} + 4O^{2-} \rightarrow Li_xMn_2O_4(s)$ $\qquad E_M(Li_xMn_2O_4)$

[The sublimation energy of lithium metal (1.65 eV), the ionisation potential of lithium (5.39 eV) and the 4$^{th}$ ionisation potential of manganese (52 eV) have been used in the steps above]

Adding all the Born steps and the corresponding energies, the net reaction and the net energy are given respectively as:

$x\,Li\,(s) + Mn_2O_4 \rightarrow Li_xMn_2O_4$ and

$$\Delta\bar{E}\,(x) = 1/8\,[x(1.65+5.39) - x\,(52) - E_M(Mn_2O_4) + E_M(Li_xMn_2O_4)$$

Now the battery voltage may be computed using equation (7):

$$V = -(1/F)\,[-5.62 + 0.125\,(dE_M/dx)] \qquad (7)$$

If the Madelung energy $E_M$ is a linear function of x, then the battery voltage will be independent of x. However in general $E_M$ may have a non-linear dependence on x in which case the battery voltage itself may depend on stoichiometry x. Despite the slight non-linearity evident in **Fig.1**, the following linear fit provides a good approximation to $E_M$.

$$E_M = 7.166\,(1.761\,x - 34.058) \qquad \text{eV / formula unit.}$$

Using equations (5) and (6), the battery voltage turns out to be 4.042 V, which agrees well with the experimental value of 4.1 V [13,14]. This is the first time that a battery voltage has been related to the Madelung energies of the electrode materials. Owing to the near linearity of **Fig. 1**, the voltage of $Li_xMn_2O_4$ is expected to depend only weakly on stoichiometry. Using quantum ab initio methods, Ceder et al have computed average intercalation voltages for layered oxide systems. If battery voltage is dependent upon x, this average method cannot capture the x dependence of the battery voltage. On the other hand, using the present method, one can compute battery voltages for any value of x.

## Conclusion

The class of ionic crystals find applications in several areas such as i) Ferro electrics ii) Piezoelectrics iii) Electrochromic devices iv) Non-linear optical materials and v) Advanced batteries and fuel cells. The ionic displacement in the crystal underlies the basic phenomenon in ferroelectrics and piezoelectrics. Hence it will be of interest to follow the electrostatic energy of the crystal as a function of ionic displacements from the normal positions. This can be implemented in our program by varying the input parameters corresponding to the ionic coordinates. Electrostatic environment in the crystal will modify the local electronic energy levels at the sites of the guest or dopant ions and hence modifying the electrochromic properties.

In batteries and fuel cells extensive material search is for suitable electrode materials. Ionic oxides of varying structures (layered, spinel) constitute an important class of electrode materials. That the Madelung energy of these materials is directly relatable to the open circuit voltages of batteries is demonstrated in this paper. A method of computing the long-range ion-ion interactions was developed in this paper for ionic crystals of variable stoichiometry and valencies. An interesting application was to battery voltage computation for $Li_xMn_2O_4$ based lithium-ion batteries where the material stoichiometry x varies continuously during battery charging and discharging. In addition to stoichiometric changes, this method will be of use in studying the effect of doping on Madelung energies. Substitution of some of the host ions by hetero ions is a widely practised way of tuning material properties for different applications.

**Appendix**

**<u>Computation of $E_{iref}$</u>**

$E_{iref}$ is the energy of interaction of any chosen reference ion with its own Bravais relatives# and with other ions in the basis and their Bravais relatives.

Let $\mathbf{r}_i = [x(i), y(i), z(i)] \qquad i = 1 \rightarrow N$

denote the atomic positions of the $i^{th}$ ion in the basis and $\lambda(i)$ $i = 1 \rightarrow N$ denote the effective charge at the $i^{th}$ ion of the basis. Shift the origin of the co-ordinates (0,0,0) so that $\mathbf{r}_{iref} = (0,0,0)$. In this co-ordinate system

$$\mathbf{r}_i\,[x(i)-x(i_{ref}),\ y(i)-y(i_{ref}), z(i)-z(i_{ref})] = \mathbf{r}_i'$$

Now the interaction energy $E_{iref}$ can be written as

$$E_{iref} = \sum_{l\neq 0} [\lambda^2(i_{ref}) / |\,\boldsymbol{l}\,|] + \sum_{i\neq i_{ref}}^{N} \sum_{l} \lambda(i_{ref})\,\lambda(i) / |\boldsymbol{l} + \mathbf{r}_i'|$$

$$= \lambda^2(i_{ref}) \sum_{l\neq 0} 1/|\boldsymbol{l}| + \lambda(i_{ref}) \sum_{i\neq iref}^{N} \lambda(i) \sum_{l} 1/|\boldsymbol{l} + \mathbf{r}_i'| \qquad \text{(i)}$$

\# Bravais relatives of a given ion are here defined as the set of ions generated by Bravais translations acting on the chosen ion.

$$E_{iref} = \lambda(i_{ref}) \Big[\Big(\sum_{l\neq 0} \lambda(i_{ref}) / |\,\boldsymbol{l}\,|\Big) + \Big(\sum_{i\neq iref}^{N} \lambda(i) \sum_{l} 1/|\boldsymbol{l} + \mathbf{r}_i'|\Big)\Big] \qquad \text{(ii)}$$

In the above equations $\boldsymbol{l}$ is the Bravais translation vector given by

$$\boldsymbol{l} = l_1\mathbf{a} + l_2\mathbf{b} + l_3\mathbf{c}$$

where the vectors **a, b** and **c** depend on the type of unit cell chosen.

Using Ewald's transformation the summations appearing in equation (ii) can be expressed as

$$\sum_{l\neq 0} 1/|\boldsymbol{l}| = \sum_{g} f(\mathbf{g}) + F(G) \qquad \text{(iii)}$$

$$\sum_{l} 1/|\boldsymbol{l} + \mathbf{r}_i'| = \sum_{g} \exp(-i\mathbf{g}.\mathbf{r}_i')\, f(\mathbf{g}) + \bar{F}(G, \mathbf{r}_i') \qquad \text{(iv)}$$

where $f(g) = (\pi/\nu_c).(1/G^2).\exp -(\mathbf{g}^2/4G^2) / (\mathbf{g}^2/4G^2)$ (v)

$$F(G) = \sum_{l\neq 0} (1/|\boldsymbol{l}|)\, \mathrm{erfc}\{G.|\boldsymbol{l}|\} - 2G/\sqrt{\pi} \qquad \text{(vi)}$$

and $\bar{F}(G, \mathbf{r}_i') = \sum_{l} (1/|\boldsymbol{l} + \mathbf{r}_i'|)\, \mathrm{erfc}\{G.|\boldsymbol{l} + \mathbf{r}_i'|\}$ (vii)

In the above equations, **G** is a variable scalar parameter which is adjusted for fast convergence of the infinite sum, **g** is the reciprocal lattice vector given by $\mathbf{g} = h\mathbf{A}+k\mathbf{B}+l\mathbf{C}$ where vectors **A, B, C** are obtained from the vectors **a, b** and **c** by the usual transformations. $\nu_c$ is the unit cell volume given by $\nu_c = |\mathbf{a} \times \mathbf{b} \bullet \mathbf{c}|$.

$E_{i\,ref}$ may now be written as

$$E_{i\,ref} = \lambda(i_{ref}).[\lambda(i_{ref}).\sum_{g} f(g) + \lambda(i_{ref}).F(G) + \sum_{g} \{\sum_{i\neq iref}^{N} \lambda(i) \exp -ig \cdot \mathbf{r}_i'\} f(g)$$

$$+ \sum_{i\neq iref}^{N} \lambda(i) \bar{F}(G, \mathbf{r}_i')] = \lambda(i_{ref}) [\sum_{g} \{\lambda(i_{ref}) + \sum_{i\neq iref}^{N} \lambda(i) \exp(-ig.\mathbf{r}_i')\} f(g)$$

$$+ F(G).\, \lambda(i_{ref}) + \sum_{i \neq iref}^{N} \lambda(i) \bar{F}\ (G, \mathbf{r}_i')] \qquad \text{(viii)}$$

The co-efficient of f(0) in the first summation appearing in the equation (viii) is

$$[\lambda(i_{ref}) + \sum_{i \neq iref}^{N} \lambda(i)] = 0$$

due to the electroneutrality of the basis. Hence the singularity arising from f ($\mathbf{g}$) for $\mathbf{g=0}$ is removed.